\documentclass[prl,floatfix,twocolumn]{revtex4}

\usepackage[dvips]{epsfig}
\usepackage{amsmath}
\usepackage{amssymb}
\usepackage{graphicx}
\usepackage{times}

\newsavebox{\CAuthor}
\begin{lrbox}{\CAuthor}
\footnotesize\verb+harmanda@mpip-mainz.mpg.de+
\end{lrbox}

\newcommand{\caplist}[1]{}

\newcommand{\romB}{{\operatorname{B}}}

\newcommand{\romc}{{\operatorname{c}}}
\newcommand{\romd}{{\operatorname{d}}}
\newcommand{\rome}{{\operatorname{e}}}
\newcommand{\romi}{{\operatorname{i}}}

\newcommand{\VECnabla}{{\boldsymbol{\nabla}}}

\newcommand{\VECq}{{\boldsymbol{q}}}
\newcommand{\VECr}{{\boldsymbol{r}}}

\newcommand{\cd}{$\cdot$}
\newcommand{\D}{}

\newcommand{\etal}{\emph{et}$\,$\emph{al.}}
\newcommand{\ie}{\emph{i.}$\,$\emph{e.}}

\renewcommand{\caplist}[1]{#1}

\begin{document}

\title{A novel method for measuring the bending rigidity of model lipid membranes by simulating tethers}

\author{Vagelis A. Harmandaris}
\author{Markus Deserno}

\affiliation{Max-Planck-Institut f\"ur Polymerforschung,
             Ackermannweg 10,
             55128 Mainz,
             Germany}

\date{\today}



\begin{abstract}
The tensile force along a cylindrical lipid bilayer tube is
proportional to the membrane's bending modulus and inversely
proportional to the tube radius.  We show that this relation, which is
experimentally exploited to measure bending rigidities, can be applied
with even greater ease in computer simulations.  Using a
coarse-grained bilayer model we efficiently obtain bending rigidities
that compare very well with complementary measurements based on an
analysis of thermal undulation modes.  We furthermore illustrate that
no deviations from simple quadratic continuum theory occur up to a
radius of curvature comparable to the bilayer thickness.
\end{abstract}

\maketitle


\newpage

\section {Introduction}

Bilayer-forming lipids are the basic structural component of
biological cell membranes. In these amphiphilic molecules a
hydrophilic group is connected to one or two hydrophobic hydrocarbon
chains. When dissolved into water they spontaneously assemble into a
variety of structures. In Nature lipid bilayers form the outer plasma
membrane of cells as well as the walls of the different cellular
compartments and organelles, such as the endoplasmic reticulum, the
Golgi apparatus, and the nucleus.
\cite{Lodish}.

Lipid bilayer membranes display interesting physics on many different
length- and time-scales.  On atomistic length scales this includes
questions such as: How do lipid tail length and its degree of
saturation influence the bilayer state, how does a specific
hydrophilic head group facilitate solubilization, or how can water
permeate the hydrophobic region?  On somewhat larger scales the
embedding of trans-membrane proteins or bilayer fusion are being
studied.  And on scales exceeding several times the bilayer thickness,
one may ask how vesicles are formed and what shape they have, which
forces go along with a particular bilayer geometry, or how the
demixing of a multicomponent membrane can trigger morphology changes.
These different sets of questions require different techniques for
their treatment.  In the present article we focus on the physics
happening on the large scale end, \ie, on the continuum level that may
be employed on length scales beyond a few tens of nanometers, when a
membrane may be viewed as a twodimensional fluid elastic sheet.

As is typical in any coarse-graining scheme, many details pertaining
the a physical system on a given scale get condensed into a few
effective parameters on a larger level.  Indeed, on the continuum
level what remains of all lipid detail are three material parameters
-- two moduli describing the softest deformation, which is bending,
and one length scale describing a spontaneous curvature.  The
respective Hamiltonian, proposed in the early 70's
\cite{Canham70,Helfrich1,Evans74} can be written as a surface integral
over the entire membrane:
\begin{equation}
\label{eq:Energy}
E = \int \romd A \, \Big\{ \frac{1}{2}\kappa(K - C_0)^2 + \bar{\kappa} K_{\text{G}} \Big\} \ .
\end{equation}
Here, the extrinsic curvature $K=1/R_1+1/R_2$ is the sum of the two
local principal curvatures, and the Gaussian curvature
$K_{\text{G}}=1/R_1R_2$ is their product.  The inverse length
$C_0$ indicates any spontaneous curvature which the bilayer might
have, so the first term quadratically penalizes deviations of the
local extrinsic curvature from $C_0$.  The two moduli $\kappa$ and
$\bar{\kappa}$ belonging to the two quadratic curvature expressions
are referred to as
\emph{bending modulus} and \emph{saddle splay modulus}, respectively.
If the membrane has two identical leaflets, $C_0=0$ by symmetry, a
situation which does seldomly hold for biological membranes but very
frequently for artificial lipid bilayers and vesicles.  Furthermore,
since the surface integral over the Ricci scalar $R$ can be expressed
as a boundary integral plus a topological term, the second term in
Eqn.~(\ref{eq:Energy}) most often only contributes a constant and can
then be ignored.  Under these conditions there remains only a single
physical parameter characterizing the membrane, the bending modulus
$\kappa$, and it is thus the most important one to determine.

Bending rigidities have been measured experimentally by various
techniques
\cite{Brochard,Webb,Bothorel,Evans1,Evans2,Ipsen1,Ipsen2,Waugh,DaiSheetz95,Evans3,Bassereau},
all ultimately based on one of two general approaches: One may either
utilize the dependence of thermal undulations on a membrane's
rigidity, or measure the force needed to actively bend it.  The
traditional realization of the first approach is to monitor the
fluctuations of vesicles as a function of wavelength by light
microscopy, a method termed ``flicker spectroscopy''
\cite{Brochard,Webb,Bothorel}.  A related experimental method is
based on micro-pipette manipulation techniques. There, the flicker
spectrum is successively suppressed by increasing the pipette
pressure, and the bending rigidity can then be obtained from the
low-tension regime of the tension-area curve
\cite{Evans1,Evans2,Ipsen1,Ipsen2}.  The second approach is typically
implemented by measuring the force needed to pull nanoscale bilayer
tubes (tethers) from vesicles \cite{Waugh,DaiSheetz95,Evans3,Bassereau}.  Since
the formation of a tube involves the creation of a high curvature, the
work to pull a tether is basically done against bending energy, hence
the modulus $\kappa$ can be determined from it.

Determination of the bending rigidity is of course equally important
in computer simulation studies of lipid bilayers, and the spectrum of
available methods is the same.  However, by far the most common
approach in simulations is flicker spectroscopy, both for atomistic
simulations \cite{LindahlEdholm00,Marrink1,BoPadeBr05} as well as for
various coarse grained methods
\cite{BoPadeBr05,Lipowsky2,Farago03,Stevens04,Ira1,Ira2,BranniganBrown06}.  Only
recently den Otter and Briels have proposed a method by which
constraining forces are applied to actively deform the membrane
\cite{Briels1}, and Farago and Pincus have proposed a scheme based on
the change in free energy of deforming the bilayer
\cite{FaragoPincus04}.  Unfortunately, both active methods involve
significant technical and conceptual sophistication.  This may explain
why the idea is not commonly employed, despite the fact that
particularly for stiff membranes fluctuation based schemes encounter
difficulties (in experiments as well as in simulations), because the
thermally excited amplitudes decrease with bending modulus and become
difficult to resolve at some point.

In the present article we propose an alternative simulation approach
for studying the curvature elasticity of membranes by an active
deformation.  Our setup essentially involves measuring the force
necessary to hold a membrane tether, and it is thus conceptually
identical to its experimental ``counterpart''.  As we will see,
complications of earlier active schemes are avoided, and the
simulations are very easy to perform and analyze.  We apply this
method to a recently proposed coarse-grained solvent-free simulation
model \cite{Ira1,Ira2} and find results that agree very well with
data from the analysis of the thermal fluctuations.
Moreover, the method permits us to check, up to
which curvatures the quadratic model from Eqn.~(\ref{eq:Energy})
remains valid.  Our results indicate that curvature radii close to the
bilayer thickness can be imposed without noticeable deviations from
Eqn.~(\ref{eq:Energy}).  While the precise location for the breakdown
of quadratic theory may well be model dependent, its validity up to
length scales comparable to bilayer thickness is in agreement with
experimental findings \cite{Bassereau}.


\section {Curvature Elasticity}

In this section we first briefly review the fluctuation approach
towards bilayer elasticity and discuss some of its difficulties.  We
then introduce the alternative scheme based on holding a membrane
tether.


\subsection {Flicker spectroscopy}

\begin{figure}
\begin{center}
\includegraphics[scale=0.765]{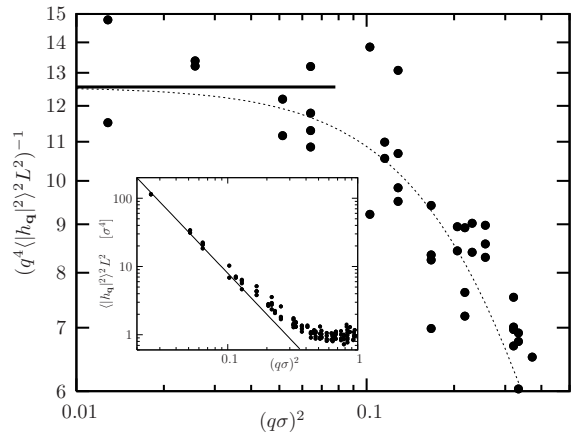}
\end{center}
\caption{
\normalsize \setlength{\baselineskip}{2em} 
Flicker spectrum of a fluctuating membrane, plotted in such a way
that in the limit $q\rightarrow 0$ the expression approaches $\kappa$,
see Eqn.~(\ref{eq:flicker}) The dashed line is a fit of the form
$(k_\romB T/\kappa+c_1\,(q\sigma)^{c_2})^{-1}$ which helps to find the
asymptotic value; the inset shows the unscaled spectrum.  The fit
leads to $\kappa=12.5\,k_\romB T$, with an error estimated to be $\pm
1\,k_\romB T$.  The system is the same lipid bilayer model
\cite{Ira1,Ira2} that will be used for the tethers, see
below.}\label{fig:flicker}
\end{figure}

The energy expression in Eqn.~(\ref{eq:Energy}) requires knowledge of
the local membrane curvature.  For essentially flat membranes, which
can be described by specifying their height $h(x,y)$ above some
reference plane (``Monge-parametrization''), this curvature is given
by
\begin{equation}
K = \VECnabla\cdot\left(\frac{\VECnabla h}{\sqrt{1+(\VECnabla h)^2}}\right)
\stackrel{|\VECnabla h|\ll
1}{\approx} \Delta h \ ,
\label{eq:K_Monge}
\end{equation}
where $\VECnabla$ is the two-dimensional nabla operator on the base
plane.  The approximation in the second step is the lowest order term
in a small gradient expansion.  On this level the Hamiltonian
(\ref{eq:Energy}) becomes quadratic and can be diagonalized by Fourier
transformation.  Assuming an $L\times L$ membrane patch with periodic
boundary conditions, and writing $h(\VECr) = \sum_{\VECq} h_{\VECq} \,
\rome^{\romi\,\VECq\cdot\VECr}$, one finds
\begin{equation}
  E = \frac{1}{2} L^2 \sum_\VECq |h_\VECq|^2 ( \kappa q^4 + \Sigma q^2 ) \ ,
\end{equation}
where we for completeness also added a surface tension term, $\Sigma$
times excess area.  From the equipartition theorem we then see that
the mean squared amplitude of each mode, \ie, the \emph{fluctuation
spectrum} or the \emph{static structure factor}, is given by
\begin{equation}
\langle  |h_\VECq|^2 \rangle = \frac{k_\romB T}{L^2\big(\kappa q^4+\Sigma q^2)} \ .
\label{eq:flicker}
\end{equation}
A fit of the fluctuation spectrum measured in the simulation to this
expression yields bending modulus $\kappa$ and tension $\Sigma$.
Since for wave vectors smaller than $q_{\min} \simeq
\sqrt{\Sigma/\kappa}$ the fluctuations are tension dominated, it is
best to simulate at zero tension in order to avoid unnecessary damping
of the most relevant modes.  In this case the expression
$1/(q^4\langle|h_\VECq|^2\rangle L^2)$ approaches $\kappa$ in the
limit $q\rightarrow 0$, as is illustrated in Fig.~\ref{fig:flicker}
for a model simulation \cite{Ira1,Ira2} described in more detail
below.  For wave vectors approaching $q_{\max}\simeq 2\pi/w$, where
$w$ is the bilayer thickness, discrete lipid fluctuations such as
protrusions require a more careful analysis \cite{BranniganBrown06}.

There are clear limitations for the calculation of $\kappa$ using
thermal fluctuations. The first and obvious one is that large values
of $\kappa$ lead to very small amplitudes. Considering ($i$) that
$\kappa/k_\romB T$ is typically of order 10 and ($ii$) how strongly
the amplitudes decay with increasing wave vector, one realizes that it
requires substantial statistics to be able to resolve the spectrum.
Particularly unpleasant in this context is that the most important
low-$q$ modes equilibrate slowest, with a time scale diverging like
$q^{-4}$.

Also, it must be appreciated that the curvatures \emph{probed} this
way are very weak.  Since (in the tensionless state) $\langle
K^2\rangle = q^4\langle |h_{\bf{q}}^{2}|\rangle =k_BT/L^2\kappa$, the
average (root-mean-square) radius of curvature is given by
$\overline{R} = \langle K^2\rangle^{-1/2} = L \sqrt{\kappa/k_\romB
T}$, \ie\ several times the box length.  These curvatures are
\emph{much} weaker than the ones which one usually imposes on systems
if one studies vesicles, budding, tubes, or fusion.  This leaves the
question open how \emph{relevant} the measured elastic constants
really are.

It is also these limitations which den Otter and Briels \cite{Briels1}
had in mind when they proposed ways for obtaining larger curvatures,
either by more advanced sampling techniques, such as umbrella
sampling, or by explicitly creating undulations with larger amplitudes
by suitable constraints.  They found that for larger amplitudes
the membrane seemed to stiffen, which might suggest that the simple
curvature elastic model underlying Eqn.~(\ref{eq:Energy}) breaks down.
However, an alternative explanation would be provided by the neglect
of higher order terms in the small gradient approximation for the
curvature in Eqn.~(\ref{eq:K_Monge}).  Indeed, for a mode $h(x,y) =
a\,\sin(qx)$ it is easy to verify that the ratio between linear and
nonlinear prediction of the curvature energy is given by
\begin{equation}
\frac{E_{\text{lin}}}{E_{\text{nonlin}}}
\stackrel{qa\gg 1}{=}
\frac{3\pi}{8}qa + \frac{3\pi}{32}(qa)^{-1} + \mathcal{O}\big((qa)^{-3}\big) \ ,
\label{eq:lin_nonlin}
\end{equation}
which diverges linearly with growing amplitude.  While this is
qualitative what den Otter and Briels observe, the magnitude of their
stiffening is bigger than what Eqn.~(\ref{eq:lin_nonlin}) would
predict.  The observed deviation for large amplitudes appears more
likely to be a result of a residual tension stemming from their
simulations being done at constant box volume.


\subsection{Stretching tethers}

Here we present a method for the calculation of $\kappa$ based on a
different approach. The basic idea is to impose a deformation of the
membrane, specifically by creating a curved cylindrical vesicle, and
then measure the force required to hold it in this deformed state.  In
the experiment such tethers are typically created by first attaching
adhesive beads to a suitably fixated giant vesicle (or a cell) and
then pulling it away with a laser tweezer that permits the measurement
of the involved force. In the simulation such a tether can simply be
stabilized by ``spanning'' a cylindrical vesicle through the
simulation box, \emph{across the periodic boundary conditions}.  One
thus simulates a system which is perfectly cylindrical (\ie, there are
no end effects), and the axial pulling force is readily obtained from
the component of the stress tensor along the box-spanning direction.

With a vesicle radius $R$ and a box length $L_z$ in the direction
of the spanned vesicle, see Fig.~\ref{fig:vesicle_snap}, the curvature energy is
\begin{equation}
\label{eq_Cyl_1}
E = \frac{\kappa }{2} \left(\frac{1}{R}\right)^2 2\pi RL_z = \frac{\pi \kappa L_z }{R} \ .
\end{equation}
The axial force under the constraint of fixed area $A=2\pi R L_z$ is
obtained from $F_z = (\partial E / \partial L_z)_{A} = 2\pi \kappa /
R$, hence the bending modulus is given by
\begin{equation}
\label{eq_Cyl_2}
\kappa = \frac{F_z R}{2\pi} \ .
\end{equation}
Since both $F_z$ and $R$ are easily measurable, $\kappa$ can be
readily determined in the simulation.  In fact, it is this point where
implementing the tether method in a simulation shows its biggest
advantage over its experimental counterpart: In a real experiment $R$
can not be measured directly, since its typical magnitude is
sub-optical. It is thus usually re-expressed in terms of the membrane
tension $\Sigma$, leading to $R=\sqrt{\kappa/2\Sigma}$ and thus
$\kappa=(F_z/2\pi)^2/2\Sigma$, but then the tension needs to be
monitored independently by other means.  Recently, however, Cuvelier
\etal\ \cite{Bassereau} devised a clever setup involving
\emph{two} tethers which avoids such complications.

Even though the above analysis is standard in the tether literature
\cite{Waugh,DaiSheetz95,Evans3,Bassereau}, it is still only approximate.
Notice that this time we have neglected thermal fluctuations
altogether.  The formula (\ref{eq_Cyl_2}) relies entirely on a
``ground state'' argument.  This is justifiable in two ways.  First,
for not too small radii of curvature the fluctuation contribution to
the force, as estimated for instance by a simple plane-wave ansatz for
the cylindrical modes, turns out to be very small.  And second, the
two most obvious effects which fluctuations have on the two terms in
Eqn.~(\ref{eq_Cyl_2}) that need to be determined, $F_z$ and $R$, are
working in opposite directions.  While clearly the mean axial force
$\langle F_z\rangle$ will increase (for exactly the same reason that
it takes a force to pull a fluctuating polymer straight), the
fluctuation-corrected mean radius $\langle R\rangle $ of the vesicle
will decrease, since the total area is constant and the area needed
for fluctuations has to come from somewhere.  Within a plane wave
approximation these two effects cancel.  A more accurate investigation
is a fair bit more subtle \cite{FournierGalatola}.

By performing various simulations of tethers with different curvature
radius $R$, we can thus address the question how far the present
quadratic theory remains valid.  Assuming symmetric membranes, the
next terms by which the Hamiltonian density in Eqn.~(\ref{eq:Energy})
needs to be amended are quartic ones, and these are $K^4$, $K^2K_{\text{G}}$,
$K_{\text{G}}^2$, and the gradient term $(\nabla_aK)(\nabla^aK)$, where
$\nabla_a$ is the metric-compatible covariant derivative
\cite{CaGuSa03}.  Since for cylinders $K_{\text{G}}=0$ and $|\nabla_a K|$=0,
the only remaining term is $K^4$.  Adding $\frac{1}{4}\kappa_4K^4$ to
the energy density and repeating the steps leading to
Eqn.~(\ref{eq_Cyl_2}), we then find
\begin{equation}
\frac{F_zR}{2\pi} = \kappa + \frac{\kappa_4}{R^2} = \kappa \, \big[1 + (\ell_4 K)^2 \big]\ ,
\label{eq_Cyl_2b}
\end{equation}
where $\ell_4=\sqrt{\kappa_4/\kappa}$ is a characteristic length scale
associated with corrections beyond quadratic order, and one typically
assumes that it is related to bilayer thickness.


\section{Mesoscopic membrane simulation}\label{sec:sim}


To illustrate our method, we have performed mesoscopic simulations of
a coarse grained lipid bilayer model recently developed in our group
\cite{Ira1}.  Roughly, lipids are represented by three consecutive
beads of diameter $\sigma$ (our unit length), with one end bead being
hydrophilic and the two tail beads hydrophobic.  The latter feature,
in addition to an excluded volume interaction, an attraction with a
tuneable depth $\epsilon$ (our unit of energy) and range $w_\romc$.
The unit of time is $\tau=\sigma\sqrt{m/\epsilon}$, where $m$ is the
unit of mass.  By properly choosing $w_\romc$ and $\epsilon$, a wide
range of self-assembling fluid bilayer phases of different bending
rigidities is obtained.  More details can be found in
Refs.~\cite{Ira1,Ira2}.

\begin{table*}
\begin{ruledtabular}
\begin{tabular}{rc|ccccccccccccc}
\# lipids & &  $L_z=20$                &   30                &   40                &   50                &   60                &   70                &   80                &   90                &   100 &  120 &  160 &  200 & 240 \\
\hline\\[-1em]
5000      & & $\D{24.0 \pm 0.5 \atop 3.96 \pm 0.9}$ & $\D{16.1 \pm 0.4  \atop 5.4 \pm 0.8}$ & $\D{12.2 \pm 0.3  \atop 6.12 \pm 0.8}$ & $\D{9.8 \pm 0.2  \atop 8.28 \pm 0.8}$ & $\D{8.3 \pm 0.2 \atop 10.44 \pm 0.8}$ & $\D{7.2 \pm 0.1 \atop 11.52 \pm 0.8}$ & $\D{6.3 \pm 0.1 \atop 12.6 \pm 0.8}$ & $\D{5.7 \pm 0.1 \atop 13.68 \pm 0.8}$ &   \cd & \cd  &  \cd &  \cd & \cd \\[1em]
10000     & & $\D{47.8 \pm 0.6 \atop 1.8 \pm 0.7}$ &  \cd & $\D{24.0 \pm 0.4  \atop 3.24 \pm 0.7}$ &  \cd & $\D{16. \pm 0.4  \atop 4.68 \pm 0.8}$ &   \cd & $\D{12.2 \pm 0.2  \atop 6.8 \pm 0.8}$ &   \cd & $\D{9.9 \pm 0.2  \atop 8.28 \pm 0.8}$ & $\D{8.3 \pm 0.1 \atop 10.5 \pm 0.7}$ & $\D{6.3 \pm 0.1 \atop 12.5 \pm 0.8}$ &  \cd & \cd \\[1em]
20000     & &  \cd &  \cd & $\D{47.7 \pm 0.6  \atop 1.7 \pm 0.8}$ &  \cd &  \cd &   \cd & $\D{24.0 \pm 0.4  \atop 3.5 \pm 0.8}$ &   \cd &   \cd & $\D{16.0 \pm 0.3  \atop 4.7 \pm 0.8}$ & $\D{12.1 \pm 0.2  \atop 6.6 \pm 0.8}$ & $\D{9.9 \pm 0.2  \atop 8.3 \pm 0.8}$ &$\D{8.3 \pm 0.1 \atop 10.4 \pm 0.8}$ \\[1em]
\end{tabular}
\end{ruledtabular}
\caption{
\normalsize \setlength{\baselineskip}{2em} 
List of all simulation geometries, sorted by the number
of lipids (first column) and their tether length (in units of
$\sigma$, first row).  In all tuples the top value indicate $\langle
R\rangle$ (in units of $\sigma$), the bottom value is $\langle
F_z\rangle$ (in units of $\epsilon/\sigma$).  We always used
$w_\romc=1.6\,\sigma$ and $k_\romB
T=1.1\,\epsilon$.}\label{tab:systems}
\end{table*}


Coarse-grained Molecular Dynamics simulations of a lipid systems with
$w_\romc=1.6\,\sigma$ and $k_\romB T=1.1\,\epsilon$ were performed
using the ESPResSo package \cite{espresso}. The geometries studied are
summarized in Table~\ref{tab:systems}. All simulations were performed
under canonical ($NVT$) conditions, using a Langevin thermostat
\cite{GrKr86} with friction constant $\Gamma= 1.0\,\tau^{-1}$ to keep
the temperature constant.  Within a rectangular box with dimensions
$L_x=L_y$ and $L_z$, using periodic boundary conditions in all
directions, a cylindrical membrane spanning the $z$-direction was
initially set up with a radius $R_{\text{setup}}$ chosen in such a way
that the area per lipid in both leaflets corresponded to the one for a
flat tensionless bilayer \cite{Ira2}.  Upon starting the simulation
$R_{\text{setup}}$ relaxed (typically within about $1000\,\tau$) to
its equilibrium value $\langle R\rangle$, which is smaller than
$R_{\text{setup}}$ by about 3-5\%\ due to the area required for
fluctuations.  For this to happen it was quite advantageous that the
flip-flop rate of lipids between the two leaflets is big enough to
permit efficient relaxation of area-difference strains going along
with changes of the mean radius.  For the integration, a time step of
$\Delta t=0.005\,\tau$ was used for most of the systems, while in some
cases we needed a smaller time step of $\Delta t=0.001\,\tau$ in order
to obtain accurate results.


\section{Results and discussion}


\begin{figure}
\begin{center}
\includegraphics[scale=1]{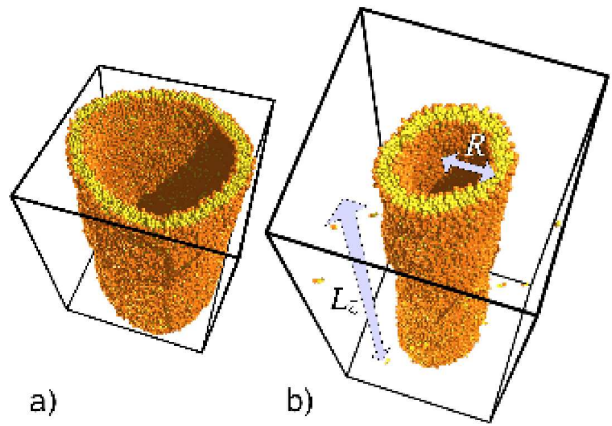}
\end{center}
\caption{
\normalsize \setlength{\baselineskip}{2em} 
Snapshots of two tether simulations with $20\,000$
lipids and different radii of curvature: a) $\langle
R\rangle=24\,\sigma$ b) $\langle
R\rangle=12\,\sigma$.}\label{fig:vesicle_snap}
\end{figure}

\begin{figure}
\begin{center}
\includegraphics[scale=0.765]{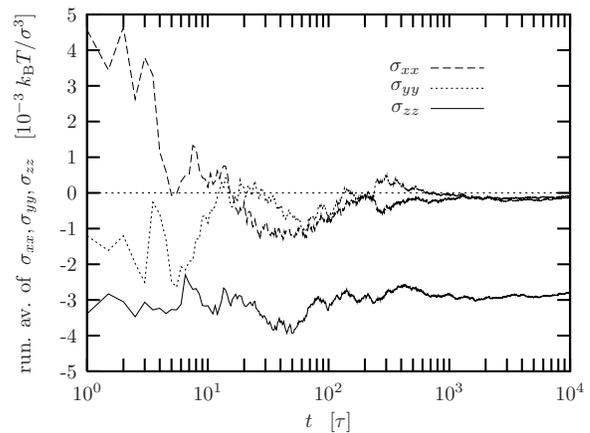}
\end{center}
\caption{
\normalsize \setlength{\baselineskip}{2em} 
Running average of the diagonal components $\sigma_{xx}$,
$\sigma_{yy}$ and $\sigma_{zz}$ of the stress tensor for a cylindrical
vesicle with $\langle R\rangle=70\,\sigma$.}
\label{fig:str_stress}
\end{figure}

Figure \ref{fig:vesicle_snap} shows two typical snapshots of
equilibrated cylindrical vesicles from different simulations.  Notice
that while fluctuations are clearly visible, they are fairly weak,
\ie, the vesicle is to a very good approximation cylindrical.  We
use the midplane between the two monolayers to denote the average
radius $\langle R\rangle$.  It is determined by first identifying the
axis, next finding the average distance of the second tail bead of the
outer leaflet to this axis, $R_{\text{out}}$, and the equivalent for
the inner leaflet, $R_{\text{in}}$.  We then take $\langle R\rangle =
\frac{1}{2}\langle R_{\text{out}}+R_{\text{in}}\rangle$, where the
average is taken during long production runs typically extending over
10000-20000 $\tau$. Errors are determined via a blocking analysis.
During these runs
we also measure the stress tensor $\sigma_{ij}$ using the virial
theorem \cite{AllenTildesley}. Figure \ref{fig:str_stress} shows a
typical example of the running average of the three normal stress
components. As we observe, the axial stress $\sigma_{zz}$ has a finite
value. In contrast, $\sigma_{xx}$ and $\sigma_{yy}$ (as well as all
off-diagonal components not shown here) approach zero.  This is
expected, since no stress is being transported across the $x$- and
$y$-direction.  The error in $\sigma_{zz}$ is also determined via a
blocking analysis.

One more point concerning the calculation of the stress tensor should
be mentioned.  In general, deriving accurately the stress (or the
pressure) from molecular dynamics simulations is not a trivial aspect.
Stress is a collective property with high statistical uncertainty
owing to the fluctuations of the instantaneous configurations. In
common simulations these fluctuations are very high due to the
relatively small size (number of particles) of the systems
studied. Therefore large systems and/or long simulation runs are
needed.  This point is particularly sever in the present case since,
as visible in Figure \ref{fig:str_stress}, we need to determine very
small values of the stress.  We have found that, in order to obtain
reliable values, the common time step used in coarse grained
simulations, $\Delta t = 0.01\,\tau$, is too long.  With this choice
$\sigma_{xx}$ and $\sigma_{yy}$ approached values which were
significantly different from zero, a clear sign of a systematic
integration error.  We thus used the smaller time step $\Delta
t=0.005\,\tau$, and for the cases of very small forces, \ie\ large
$\langle R\rangle$, even a time step of $\Delta t=0.001\,\tau$.

\begin{figure}
\begin{center}
\includegraphics[scale=0.765]{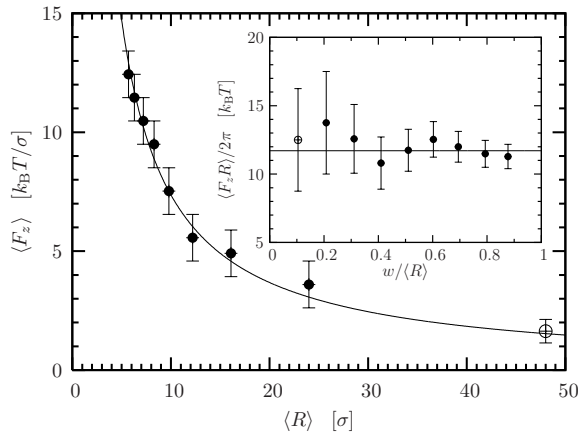}
\end{center}
\caption{
\normalsize \setlength{\baselineskip}{2em} 
Tensile force $\langle F_z\rangle$ as a function of cylinder
radius $\langle R\rangle$ for the systems with $N=5000$ lipids (with
the exception of the open symbol, which has $N=10\,000$ lipids).  The
solid line is a fit to Eqn.~(\ref{eq_Cyl_2}), leading to
$\kappa=11.7\,k_\romB T$.  The inset shows the combination $F_z
R/2\pi$ as a function of rescaled curvature $w/\langle R\rangle$, and
the solid line again indicates $\kappa=11.7\,k_\romB T$.}
\label{fig:str_force}
\end{figure}

The tensile force in $z$ direction is obtained from the stress via
$F_z=\sigma_{zz}L_xL_y$.  Figure \ref{fig:str_force} shows the force
for the systems with $N=5000$ lipids as a function of the average
cylinder radius $\langle R\rangle$.  As the radius increases, the
tensile force $F_z$ decreases in accord with Eqn.~(\ref{eq_Cyl_2}).
This is seen even better by looking at the combination $F_z R/2\pi$,
which is shown in the inset of Fig.~\ref{fig:str_force} as a function
of rescaled curvature $w/\langle R\rangle$.  Notice that within the
error bars of our simulation this expression is perfectly compatible
with a constant; a fit gives $\kappa/k_\romB T=11.7\pm 0.2$, in very
good agreement with the value $\kappa/k_\romB T=12.5\pm 1$ obtained
from an analysis of thermal undulation modes (see
Fig.~\ref{fig:flicker}).  A possible quadratic deviation, as suggested
by Eqn.~(\ref{eq_Cyl_2b}), can not be identified with any statistical
significance.  This is all the more amazing when we see that the most
strongly curved cylinder has $w/R\approx 0.9$, \ie, a radius of
curvature which is only 10\%\ larger than the bilayer thickness $w$!
Stated differently, the length $\ell_4$ from Eqn.~(\ref{eq_Cyl_2b})
must be a fair bit smaller than $w$.  This remarkable robustness of
the simple quadratic Helfrich theory down to such small radii of
curvature might of course be a special feature of the particular model
we have studied, and it would be worthwhile to subject other
coarse-grained lipid models to a similar test.  But the fact that
extremely high curvatures can be imposed without noticing deviations
from quadratic continuum theory is in accord with common practice in
tether pulling experiments, where the radii of curvature of these
membrane tubes are typically in the 10-40$\,\text{nm}$ range,
apparently without ever having triggered the need to include higher
order corrections to the elastic behavior
\cite{Bassereau}.

Another practical aspect of our proposed method, is related to
numerical efficiency.  The traditional method of analyzing the thermal
fluctuation spectrum requires very long simulation runs, since ($i$)
large systems need to be studied in order to have a series of wave
vectors in the regime where continuum methods are applicable, and
($ii$) these long wavelength modes take a particularly long time to
equilibrate (the relaxation time of bending modes scales with the
fourth power of wavelength).  Strictly speaking our method also
requires large systems to be studied in order to extrapolate to the
zero curvature limit (\ie, $R\rightarrow\infty)$.  However, as we have
seen in Fig.~\ref{fig:str_force}, the asymptotic limit in our case is
already reached for fairly small systems which still have a
significant curvature.  If the same holds for other lipid models -- a
fact that needs to be checked -- their value of $\kappa$ can also be
determined via the tether method using fairly small systems and
corresponding small simulation times.  For example, the point in
Figure \ref{fig:str_force} having $w/\langle R\rangle=0.5$ is taken
from a run of only 2-3 days on a single AMD Opteron 2.2GHz
processor. For the same system the analysis of the thermal
fluctuations needs at least one month on the same machine.

\begin{figure}
\begin{center}
\includegraphics[scale=0.765]{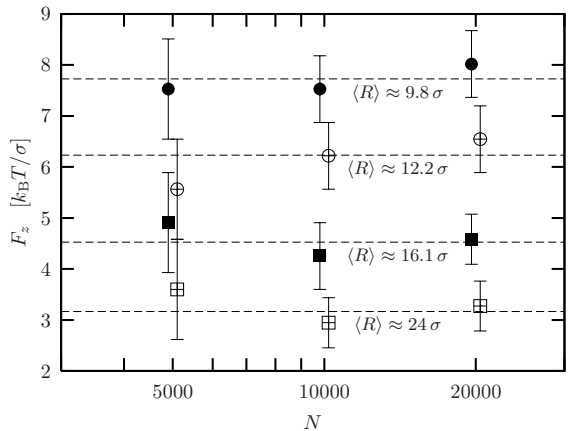}
\end{center}
\caption{
\normalsize \setlength{\baselineskip}{2em} 
Tensile force $F_z$ as a function of system size for systems
of different ratio $L_z/N$: {\Large $\bullet$}: 0.01, {\Large
$\circ$}: 0.008, $\blacksquare$: 0.006, $\square$: 0.004.  The
corresponding average radius $\langle R\rangle$ is als indicated.  A
slight horizontal offset is included to improve visibility of the
error bars.}
\label{fig:str_force_All}
\end{figure}

As a final point we would like to address the issue of finite size
effects.  While the tether force obviously depends on its radius,
Eqn.~(\ref{eq_Cyl_2}) suggests that tether \emph{length} $L_z$ (or
equivalently, the number $N$ of lipids) is irrelevant.  This is indeed
rigorously true in the ground state, but fluctuations might change the
picture.  We have thus repeated our simulations for systems with a
different number of lipids and checked, whether systems with a fixed
ratio $L_z/N$ (\ie, essentially identical radius) but varying $N$ show
any noticeable systematic change in the tensile force $F_z$.
Figure~\ref{fig:str_force_All} shows the results of such simulations.
As can be seen, the measured forces are, at least within our error
bars, compatible with a constant value for a fixed ratio $L_z/N$.  No
finite size effect is detectable.  Notice that this also provides
another independent check that fluctuations in our system, even though
present, are a subdominant effect compared to the main ``signal''
which is well described by ground state theory.

\section {Conclusions}


We have presented a new method for calculating the bending rigidity of
lipid membranes in simulations.  It involves the simulation of
cylindrical membrane tethers, spanned across the periodic boundary
conditions of the simulation box, and measuring their equilibrium
radius as well as the tensile force they exercise on the box.  In
contrast to fluctuation based schemes, which monitor thermally excited
shape deformations, our approach actively imposes a deformation on the
system and measures the restoring force and is thus not limited to the
regime of deformations accessible by thermal energy.  In fact, thermal
undulations only contribute a small correction to the main observable,
in stark contrast to fluctuation schemes in which they provide the
dominant signal.  For this reason our method is very efficient, also
applicable to stiff membranes which show very small undulations to
begin with, and does not crucially depend on the relaxation of very
slow long wavelength modes.  The straightforward access to strong
bending permits a check of quadratic continuum theory, without running
into difficulties of Monge gauge and its linearization.  For the
coarse-grained lipid model we explicitly studied we showed continuum
theory to be applicable up to curvatures comparable to bilayer
thickness.  Finite size effects would originate from fluctuations and
are thus also weak; in our runs they were not detectable.

We believe that this method provides a powerful alternative to the
existing schemes that is worth to be applied to other existing
coarse-grained models.  Not only is an independent measurement of the
elastic modulus very valuable, determining the range of validity of
continuum theory for each model would be an important bit of
knowledge, given that the curvatures that are regularly imposed in
simulations exceed thermally excited ones by at least one or two
orders of magnitude.

\section*{Acknowledgements}

We are grateful to I. Cooke for his help with the coarse-grained
simulations, and to J.-B. Fournier, W. den Otter, W. Briels,
B. Reynolds, N. van der Vegt and K. Kremer for useful discussions.  MD
acknowledges financial support by the German Science Foundation under
grant De775/1-3.


\end{document}